\documentclass[preprint,eqsecnum,nofootinbib,floats,aps]{revtex4}
\usepackage{graphicx}
\usepackage{bm}
\usepackage{epsfig}
\begin{document}
\setlength{\textheight}{9.5 in}
\newcommand{\br}{\begin{eqnarray}}
\newcommand{\er}{\end{eqnarray}}
\newcommand{\ket}[1]{\mid{#1}\rangle}
\newcommand{\go}{\longrightarrow}
\newcommand{\gsim}{\stackrel{\sim}{>}}
\def\O {{{\cal O}}}
\newcommand{\M}{\cal{M}}
\def\Wgo {\stackrel{\mbox{\tiny $FNS$ }}{\longrightarrow}}
\def\mbn{\mbox{\boldmath$\nabla$}}
\def\intsum {\int\!\!\!\!\!\!\!{\ss \sum}}
\def\intsums {\int\!\!\!\!\!{\ss\Sigma}}
\def\rbh {\hat{\bf r}}
\def\Lb {{\bf L}}
\def\Xb {{\bf X}}
\def\Yb {{\bf Y}}
\def\Zb {{\bf Z}}
\def\rb {{\bf r}}
\def\kb {{\bf k }}
\def\qb {{\bf q}}
\def\kr {{\bf k}\cdot{\bf r}}
\def\pb {{\bf p}}
\def\etal{{\it et al. }}
\def\cf{{\it cf }}
\def\ie{{\em i.e. }}
\def\etc{ {\it etc}}
\def\dket#1{||#1 \rangle}
\def\dbra#1{\langle #1||}
\def\isim{\:\raisebox{-0.5ex}{$\stackrel{\textstyle.}{=}$}\:}
\def\lsim{\:\raisebox{-0.5ex}{$\stackrel{\textstyle<}{\sim}$}\:}
\def\gsim{\:\raisebox{-0.5ex}{$\stackrel{\textstyle>}{\sim}$}\:}
\def\a {{\alpha}}
\def\b {{\beta}}
\def\e {{\epsilon}}
\def\g {{\gamma}}
\def\r {{\rho}}
\def\s {{\sigma}}
\def\k {{\kappa}}
\def\l {{\lambda}}
\def\m {{\mu}}
\def\n {{\nu}}
\def\r {{\rho}}
\def\t {{\tau}}
\def\w {{\omega}}
\def\be{\begin{equation}}
\def\ee{\end{equation}}
\def\br{\begin{eqnarray}}
\def\er{\end{eqnarray}}
\def\brn{\begin{eqnarray*}}
\def\ern{\end{eqnarray*}}
\def\x{\times}
\def\go{\rightarrow  }
\def\rf#1{{(\ref{#1})}}
\def\nn{\nonumber }
\def\ket#1{|#1 \rangle}
\def\bra#1{\langle #1|}
\def\Ket#1{||#1 \rangle}
\def\Bra#1{\langle #1||}
\def\ov#1#2{\langle #1 | #2  \rangle }
\def\hf {{1\over 2}}
\def\hw {\hbar \omega}
\def\mbs{\mbox{\boldmath$\sigma$}}
\def\gsim{\:\raisebox{-0.5ex}{$\stackrel{\textstyle>}{\sim}$}\:}
\def\lsim{\:\raisebox{-0.5ex}{$\stackrel{\textstyle<}{\sim}$}\:}
\def\mbn{\mbox{\boldmath$\nabla$}}
\def\sss{\scriptscriptstyle}
\def\ss{\scriptstyle}
\def\endauthors{}
\def\authors#1\endauthors{#1}
\def\ex#1{\langle #1 \rangle }
\def\ninj#1#2#3#4#5#6#7#8#9{\left\{\negthinspace\begin{array}{ccc}
#1&#2&#3\\#4&#5&#6\\#7&#8&#9\end{array}\right\}}
\def\sixj#1#2#3#4#5#6{\left\{\negthinspace\begin{array}{ccc}
#1&#2&#3\\#4&#5&#6\end{array}\right\}}
\def\threej#1#2#3#4#5#6{\left(\negthinspace\begin{array}{ccc}
#1&#2&#3\\#4&#5&#6\end{array}\right)}
\def\sixja#1#2#3#4#5#6{\left\{\negthinspace\begin{array}{ccc}
#1&#2&#3\\#4&#5&#6\end{array}\right\}}
\def\bin#1#2{\left(\negthinspace\begin{array}{c}#1\\#2\end{array}\right)}
\def\ul{\underline}
\def\ol{\overline}
\def\kh {\hat{k}}
\def\rh {\hat{r}}
\def\bit{\begin{itemize}}
\def\eit{\end{itemize}}
\def\bnu{\begin{enumerate}}
\def\enu{\end{enumerate}}
\def\ropp{\rho_{p'}}
\def\ronp{\rho_{n'}}
\def\rop{\rho_{p}}
\def\ron{\rho_{n}}
\def\ubpp{\bar{u}_{{p}'}}
\def\vbpp{\bar{v}_{{p}'}}
\def\ubnp{\bar{u}_{{ n}'}}
\def\vbnp{\bar{v}_{{ n}'}}
\def\upp{u_{{ p}'}}
\def\vpp{v_{{ p}'}}
\def\unp{u_{{ n}'}}
\def\vnp{v_{{n}'}}
\def\ubp{\bar{u}_{{ p}}}
\def\vbp{\bar{v}_{{ p}}}
\def\ubn{\bar{u}_{{ n}}}
\def\vbn{\bar{v}_{{ n}}}
\def\l {{\lambda}}
\def\fot{\frac{1}{2}}
\def\slash#1{\not\!{#1}}
\def\eabc{\epsilon^{abc}}
\def\rhn {\hat{\rb}_n}
\def\rhm {\hat{\rb}_m}
\def\khb {\hat{\kb}}
\def\rbh{\hat{\rb}}
\def\up{u_{p}}
\def\vp{v_{p}}
\def\un{u_{n}}
\def\vn{v_{n}}
\def\upp{u_{p'}}
\def\vpp{v_{p'}}
\def\unp{u_{n'}}
\def\vnp{v_{n'}}
\def\betago {\stackrel{\mbox{\tiny\, $\beta^-$ }}{\rightarrow}}
\def\betapgo {\stackrel{\mbox{\tiny\, $\beta^+$ }}{\rightarrow}}
\def\betabetago {\stackrel{\mbox{\tiny\,\, $\beta\beta^-$ }}
{\longrightarrow}}

\vspace{2cm}
\title{ A Novel Nuclear Model for  Double Beta Decay }
\author{ Franjo Krmpoti\'c}
\affiliation{Instituto de F\'{\i}sica, Universidade de S\~ao Paulo,
05315-970 S\~ao Paulo-SP, Brazil}
\affiliation{Instituto de F\'{\i}sica La Plata,
CONICET, 1900 La Plata, Argentina}
\affiliation{Facultad de Ciencias Astron\'omicas y Geof\'{\i}sicas,
Universidad Nacional de La Plata, 1900 La Plata, Argentina,}

\date{\today}
\begin{abstract}
The possibility of applying the Quasiparticle Tamm-Dancoff Approximation (QTDA) to describe the nuclear double beta
decay is explored. Several serious inconveniences found in the  Quasiparticle Random Phase Approximation (QRPA),
such as:
 i) the extreme sensitivity of the $2\nu\beta \beta$ decay amplitudes ${\cal M}_{2{\nu}}$ on the residual
 interaction in the particle-particle channel,
 ii) the ambiguity in treating the intermediate states, and
 iii) the need for performing a second charge-conserving QRPA to describe the $\beta\beta$-decays to the excited
 final states, are not present in the QTDA. Also, the QTDA allows for explicit evaluation of energy distributions
 of the double-charge-exchange transition strengths and of their sum rules, and can be straightforwardly
 applied to single- and double-closed shell nuclei. As an example, the $^{48}$Ca$\go^{48}$Ti decay is discussed
 within the $1fp$-shell in the particle-hole limit of the QTDA. The general [$(1,1)$-Pad\'e-approximant-like]
 behavior of the $2\nu\beta \beta$-decay amplitude in the plain QRPA as well as within its different variations
 is briefly reviewed.
\end{abstract}

 \maketitle

\maketitle
\newpage
\section{Introduction}
It is a great pleasure and a great honor for me to contribute to this commemorative issue in memory of Dubravko
Tadi\'c, who was one of my closest friends and the best coworker I ever had. I miss him badly, as many people do!
 We cooperated closely since 1966. First, we studied the single beta ($\beta$)-decay \cite{Krm66,Ema67,Krm69,Ema72},
  and in recent years we  were basically involved in the double beta ($\beta\beta$)-decay
  \cite{Bar96,Bar97,Bar98,Bar99}, and the nonmesonic weak decay of hypernuclei
  \cite{ Bar01,Bar01a,Bar01b,Bar02,Krm03}. These topics are nice examples of interrelation between Particle and
  Nuclear Physics. In fact, Dubravko Tadi\'c took part in many important developments of the theory of weak
  interactions, as well as in the advancement of particle and nuclear physics as a whole. With the entanglement
  between birds and fishes in the Escher's engraving, shown in Figure 1, I want to symbolize the close
  cooperation I had with Dubravko. To tell the truth, I have done my first work in theoretical physics,
  entitled: {\em On the Induced Terms and Partial Conservation of the Axial Vector Current in Beta Decay}
  \cite{Krm66}, under Dubravko's guidance, and the line of research in our last common work, entitled:
  {\em Nuclear Structure in Nonmesonic Weak Decay of Hypernuclei} \cite{Krm03}, has also been suggested by
  Dubravko. Therefore it is not difficult to figure out who was the bird and who was the fish in our teamwork.

\begin{figure}[h]
\begin{center}
   \leavevmode
   \epsfxsize = 10cm
     \epsfysize = 9cm
    \epsffile{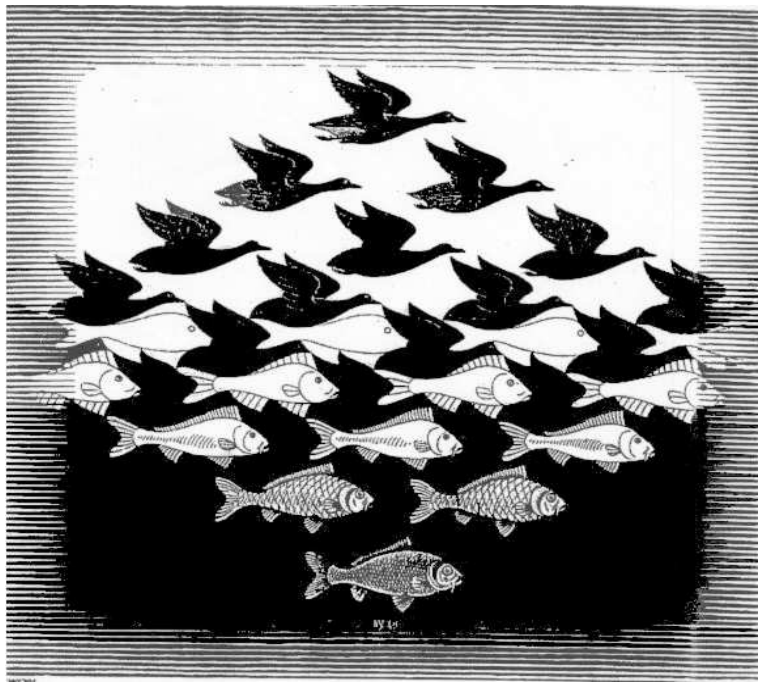}
\end{center}
\caption{\footnotesize
Escher's engraving  where  the entanglement between birds and fishes
pictures my joint work  with Dubravko.}
\label{fig1}\end{figure}

Among several topics on weak interactions that I have tackled with Dubravko, I will limit the present discussion to
a few features of the $\beta\beta$-decay, from which we can learn about the neutrino physics, provided we know
how to deal with the nuclear structure. This is a second-order weak process whose electromagnetic analogies
are the atomic Raman scattering and the  nuclear $\gamma \gamma$-decay \cite{Kra87}.

In nature there are about 50 nuclear systems where the single
 $\beta$-decay is  energetically forbidden,  and therefore the $\beta\beta$-decay turns out to be the only
 possible mode of disintegration.
It is the nuclear pairing force which causes such an "anomaly", by making the mass of
the odd-odd isobar, $(N - 1,Z + 1)$, to be greater than the masses of its even-even neighbors,
$(N,Z)$ and $(N ,- 2,Z + 2)$.

The modes by which the $\beta \beta$-decay can take place are
connected with the
neutrino ($\nu$)-antineutrino ($\tilde{\nu}$) distinction. If
$\nu$ and $\tilde{\nu}$ are defined by the transitions:
\br
&&n \go p+e^-+\tilde{\nu}\\
\nonumber
&&\nu+n\go p+e^-,
\label{1.1}\er
 the two-neutrino double beta ($2\nu\beta \beta)$
decay $(N,Z) \betabetago (N-2,Z+2)$ can occur by
two successive $\beta$ decays:
\br
(N,Z)&\betago &(N-1,Z+1)+e^-+\tilde{\nu}
\nonumber\\
&\betago &(N-2,Z+2)+2e^-+2\tilde{\nu},
\label{1.2}\er
passing through the intermediate virtual states of the $(N-1, Z+1)$ nucleus.

  However, neutrino is the only fermion that lacks a
conserved  additive quantum number to differentiate between $\nu$ and
$\tilde{\nu}$.
Thus, it is possible for the neutrino to be a Majorana particle
($\tilde{\nu}=\nu $), \ie equal to its own antiparticle
\`{a} la $\pi^0$ \footnote {A Dirac particle can be viewed as a
combination of two Majorana particles with equal masses and opposite
CP properties, in which case their contributions to the $0\nu\beta
\beta$ decay cancel. }. In this case  the
neutrinoless $\beta \beta$   ($0\nu\beta \beta$) decay is also allowed:
\br
(N,Z)&\betago &(N-1,Z+1)+e^-+\tilde{\nu} \equiv (N-1,Z+1)+e^-+\nu,
\nonumber\\
&\betago &(N-2,Z+2)+2e^-.
\label{1.3}
\er
In absence of helicity suppression (as would be natural before the
parity violation has been observed) this $0\nu$
mode is favoured over the $2\nu$
mode by a phase-space factor of $10^7-10^9$: $T_{0\nu}\sim (10^{13}-10^{15})$
yr, while $T_{2\nu}\sim (10^{20}-10^{24})$ yr.

By the early 1950's several searches for the $\beta \beta$ decay were performed, inferring  that
$T_{2\nu+0\nu} \gtrsim 10^{17}$ yr. This pointed towards $\tilde{\nu}\neq \nu$, and prompted the
introduction of the lepton number $L$ to distinguish $\nu$ from $\tilde{\nu}$: $L=+1$ was
attributed to $e^-$ and $\nu$ and $L=-1$ to $e^+$ and $\tilde{\nu}$.
The assumption on conserving the additive lepton number  allows
the $2\nu\beta \beta$ decay but forbids the $0\nu\beta \beta$ one,
for which $\Delta L=2$.

    But in 1957, with the discovery of parity non-conservation in weak interactions, the question of the
     Majorana/Dirac character for the neutrino is raised again.
That is, the neutrino was found
to be left-handed (LH) and the  antineutrino right-handed (RH), and in
place of \rf{1.1} one now has:
\br
&&n \go p+e^-+{\nu}_{RH},\\
\nonumber
&&{\nu}_{LH}+n\go p+e^-.
\label{1.4}\er
Consequently the second process in (\ref{1.3}) is forbidden because the right-handed neutrino, emitted in the
first step, has the wrong helicity to be reabsorbed in the second step. For massless neutrinos, as was
believed they were, there is no mixture of ${\nu}_{LH}$ in ${\nu}_{RH}$, and the $0\nu\beta \beta$-decay
cannot go through, regardless of Dirac or Majorana nature of the neutrino. This event discouraged experimental
work for a long time.

However, with the development of modern gauge theories the state of affairs began to change, and over the past
decade the interest in the $\beta \beta$-decay  sprang up again. There are many reasons for that.
The most important one is the fact that the $0\nu\beta \beta$-decay plays a decisive role in shaping the
ultimate theory in any new physics beyond the standard $SU(2)_L\times U(1)$ gauge model of electroweak
interactions. Moreover, no solid theoretical principle prevents neutrinos from having mass, while the most
attractive extensions of the standard model require neutrinos to be massive. Nor does theory predict the
scale of neutrino masses any better than it can fix the masses of quarks and charged leptons.

Yet, once the neutrino becomes massive, the  helicity is not a good quantum number any more.
Then, if the neutrino is in addition a Majorana particle with an effective mass $\langle m_\nu \rangle$,
the mixture of ${\nu}_{LH}$ in ${\nu}_{RH}$ is proportional to $\langle m_\nu \rangle/ E_\nu$,
and  $0\nu\beta \beta$-decay is allowed
 \footnote{For simplicity we assume that weak interactions with
right-handed currents do not play an essential role in the
neutrinoless decay.}.
This fact inspired experimental searches in many nuclei,
not only for the ${0\nu}$ $\beta \beta$-decay but also for the ${2\nu}$-decay, since these two modes of
disintegration are related through  the
  the nuclear structure effects.
In fact, their half-lives  cast in
the form:
\br
T_{2\nu}^{-1}={\cal G}_{2\nu}{\cal M}_{2\nu}^2,
\hspace{2cm}
T_{0\nu}^{-1}={\cal G}_{0\nu}{\cal M}_{0\nu}^2 \langle m_\nu  \rangle^2,
\label{1.5}\er
where ${\cal G}'s$ are  geometrical phase space
factors, and  ${\cal M}'s$ are
nuclear matrix elements (NME's).  ${\cal M}_{2\nu}$ and  ${\cal M}_{0\nu}$
present many similar features to the extent that it can be stated that we shall not understand the
$0\nu\beta \beta$-decay unless we understand the $2\nu\beta \beta$-decay.

I will limit the discussion here to the $2\nu$ mode, but the whole presentation that follows can be
straightforwardly applied to the $0\nu$ mode. With Dubravko we developed a full formulation of the
$0\nu\beta \beta$-decay, based on the Fourier-Bessel expansion of the weak Hamiltonian,  expressly
adapted for nuclear structure calculations \cite{Bar98, Bar99}. We have also worked  together on the
``charged majoron models'' \cite{Bar96,Bar97}, so called because the majoron carries the unbroken
$U(1)$ charge of lepton number.
    These models are probably the only ones that have a chance of producing the
    neutrinoless $\beta \beta$-decay that includes the emission of a massless majoron at a rate which
    could be observed in the present generation of experiments.

At present we have at our disposal beautiful data on several $2\nu\beta \beta$-decays, namely in the
following nuclei:  $^{48}$Ca, $^{76}$Ge, $^{82}$Se, $^{96}$Zr, $^{100}$Mo, $^{116}$Cd, $^{128}$Te, $^{130}$Te,
$^{136}$Xe, $^{150}$Nd, $^{238}$U and $^{244}$Pu. The $2\nu\beta \beta$-decay turned out to be one of the
slowest processes observed so far in nature and offers a unique opportunity for testing the nuclear
physics techniques for half-lives $ \gtrsim 10^{20}$ yr.
 Disappointingly, as yet, after many years of heroic efforts of many physicists, no evidence of the
 nonstandard $\beta\beta$-decay has appeared.
A survey of experimental results is given in the review article by Ejiri \cite{Eji00}.

All we know is that the massive neutrinos can seesaw. In fact, the major advances in neutrino massiveness
have been based so far on the compelling evidence that favors the neutrino oscillations. They first
came from atmospheric neutrino flux measurements at SuperKamiokande (SK) \cite{sk98}, and the  solar
neutrino shortages at SAGE, Gallex, GNO \cite{sol}, Kamiokande and SK \cite{SKsolar}. However, only
recently,  the solar SNO experiment \cite{sno1}, jointly with the reactor KamLAND (KL) experiment
\cite{kl162}, and the first long baseline accelerator K2K experiment \cite{k2k}, firmly fixed the neutrino oscillations.

That is, we now know the neutrino mix, and we have initial values for their mixing matrix elements.
 We know the number of light active neutrino species and the differences between the squares of their masses.
But we still don't know two crucial features of the neutrino
physics, which are:
 a) the absolute mass scale, and b) whether
the neutrino is a Majorana or a Dirac particle.
Only  the $0\nu\beta \beta$-decay  can  provide this information,
and this fact has motivated the undertaking of very
attractive next generation  experiments for many different isotopes,
including $^{48}$Ca, $^{76}$Ge,
$^{100}$Mo, $^{116}$Cd, $^{130}$Te, $^{136}$Xe,
$^{150}$Nd, and $^{160}$Gd \cite{Bah04,Cra04}.

The interest in neutrinos goes beyond the study of their
intrinsic properties, and extends to a variety of topics in
astro-nuclear physics, such as
the understanding of the energy production in our sun, the synthesis
of heavy elements during the
r-process, the influence of neutrinos on the dynamics of a
core-collapse supernova explosion and the cooling of a
proto-neutronstar.
The neutrino physics  even appears in cosmological questions such
as the role of neutrinos in the matter-antimatter asymmetry in the
universe.

The outline of this paper is as follows:
In Section II we list  a few general features of the NME's, which are necessary for understanding what follows. We discuss  in Section III
the general behavior  of the $\beta\beta$-decay amplitude within the
charge-exchange QRPA,  and we record the matching refinements as well.
We develop in Section IV a simple nuclear model for the $\beta\beta$-decay, based on the
well-known Quasiparticle Tamm-Dancoff  Approximation (QTDA).
In Section V we do the analysis of the
 $^{48}$Ca$\betabetago {^{48}}$Ti $2\nu$-decay   in the particle-hole
limit of this model.  A few final comments and remarks are pointed out in Section VI.

\section{${2\nu}\beta \beta$ matrix element}

Independently of the nuclear model used, and when only allowed transitions are considered,
the ${2\nu}\beta \beta$ matrix element for the $\ket{0^{+}_f}$ final state reads
\begin{eqnarray}
{\cal M}_{2{\nu}}(f)&=&\sum_{\lambda=0,1}(-)^\lambda
  \sum_{\alpha} \left[\frac
{\Bra{0^{+}_f}\O^{\beta^-}_\lambda\Ket{\lambda^{+}_\alpha}
\Bra{\lambda^{+}_\alpha}\O^{\beta^-}_\lambda\Ket{0^{+}}}
{{\cal D}_{\lambda ^+_\alpha,f}} \right]
\equiv {\cal M}_{2{\nu}}^{F}(f)+ {\cal M}_{2\nu}^{GT}(f)
\label{2.1}\end{eqnarray}
where  the summation goes over all intermediate virtual states
$ \ket{ \lambda^{+}_\alpha} $,
\br
&&\O^{\beta^-}_\lambda=(2\lambda+1)^{-1/2}\sum_{pn}\Bra{p}{\rm O}_\lambda\Ket{n}
\left(c^{{\dagger}}_p c_{\bar{n}}\right)_\lambda,
\hspace{1cm}\mbox{with}\hspace{1cm} \left\{\begin{array}{ll}
{\rm O}_0=1\;\;& \mbox{for F} \;\\
{\rm O}_1=\sigma\;\;& \mbox{for GT} \;\\
\end{array}\right.
\label{2.2}
\end{eqnarray}
are the Fermi (F) and Gamow-Teller (GT) operators for
$\beta^-$-decay, and $c^{\dagger}$ ($c$) are the particle creation
(annihilation) operators.
The corresponding $\beta^+$-decay operators are
$\O^{\beta^+}_\lambda=\left(\O^{\beta^-}_\lambda\right)^\dagger$, and
\begin{eqnarray}
{\cal D}_{\lambda^{+}_\alpha,f }
={E}_{\lambda^{+}_\alpha }-\frac{E_0+E_{0^+_f}}{2}=
{E}_{\lambda^{+}_\alpha }-E_0-\frac{E_{0^+_f}-E_0}{2},
\label{2.3}\end{eqnarray}
is the energy denominator. $E_0$ and  $E_{0^+_f}$ are, respectively, the
energy  of the initial state $\ket{0^+}$  and
of the final states $\ket{0^+_f}$.

Contributions from the first-forbidden operators, which appear in the multipole expansion of the weak
Hamiltonian, as well as those from the weak-magnetism term and other second order corrections on the
allowed $2\nu\beta \beta$-decay are not relevant for the present work and will not been tackled here.
 In recent years we have examined all of them rather thoroughly \cite{Bar95, Bar99}.

In nuclear physics the isospin symmetry is conserved to a great extent, while the Wigner SU(4)
symmetry is not.  Because of this, the amplitude ${\cal M}_{2{\nu}}^{F}$ is often neglected in
the literature. Nevertheless, one should keep in mind that, while the mean field strongly breaks
the isospin symmetry, the residual force restores it almost fully. Therefore, although in many
cases the final value of ${\cal M}_{2{\nu}}^{F}$ is small,   it is recommendable to keep track of
this NME during
calculation so as to test the consistence of the nuclear model, as well as to fix its coupling constants.

The energy distributions of the transition strengths
$|\Bra{\lambda_\a^+}\O^{\beta^{\pm}}_\lambda\Ket{0^+}|^2$ link
the single $\beta^{\pm}$-decays
to the single charge-exchange reactions, such as $(p, n)$, $(n, p)$, \etc~
\cite{Eji00,Ike65,Ebe75,Krm81,Nak82,Hir90}.
The total $\beta^{\pm}$ strengths
\begin{equation}
 S^{\beta^{\pm}}_\lambda =(2\lambda+1)^{-1}
\sum_{\a}|\Bra{\lambda_\a^+}\O^{\beta^{\pm}}_\lambda\Ket{0^+}|^2,
\label{2.4}
\end{equation}
can be expressed in the form
\begin{equation}
 S^{\beta^{\pm}}_\lambda=
\bra{0^+}\O^{\beta^{\mp}}_\lambda\cdot
\O^{\beta^{\pm}}_\lambda\ket{0^+}\equiv(-)^\lambda
(2\lambda+1)^{-1}\bra{0^+}[\O^{\beta^{\mp}}_\lambda
\O^{\beta^{\pm}}_\lambda]_0\ket{0^+},
\label{2.5}
\end{equation}
when $\ket{\lambda_\a^+}$ is a complete set of excited states
that can be reached by acting with $\O^{\beta^{\mp}}_\lambda$ on
the initial
state $\ket{0^+}$.
 It follows at once that
\begin{equation}
S_\lambda^\beta\equiv S^{\beta^{-}}_\lambda-S^{\beta^{+}}_\lambda
=(-)^\lambda
(2\lambda+1)^{-1}\bra{0^+}[\O^{\beta^{+}}_\lambda,
\O^{\beta^{-}}_\lambda]_0\ket{0^+}=N-Z,
\label{2.6}
\end{equation}
which is the well-known single-charge-exchange sum rule, also
called Ikeda sum rule (ISR) \cite{Ike65}, for both the F and the GT transitions.

Similarly,
the  $\beta\beta$-decays are closely related to the
double-charge-exchange reactions, and  to the spectral
distribution of their  strengths,
\br
 S^{\beta\beta^{\pm}}_\lambda(f)
&=&(2\lambda+1)^{-1}
|\sum_{\a}\Bra{0^{+}_f}\O^{\beta^{\pm}}_\lambda\Ket{\lambda_\a^+}
\Bra{\lambda_\a^+}\O^{\beta^{\pm}}_\lambda
\ket{0^{+}}|^2,
\label{2.7}\end{eqnarray}
over the final states $\ket{0^{+}_f}$.
The total strengths are defined as:
\br
 S^{\beta\beta^{\pm}}_\lambda&=&\sum_fS^{\beta\beta^{\pm}}_\lambda(f)
=(2\lambda+1)^{-1}
\sum_f|\bra{0^{+}_f}\O^{\beta^{\pm}}_\lambda\cdot\O^{\beta^{\pm}}_\lambda
\ket{0^{+}}|^2
\label{2.8}\end{eqnarray}
and can be rewritten in the form
\br
 S^{\beta\beta^{\pm}}_\lambda
=(2\lambda+1)^{-1}
\bra{0^{+}}\O^{\beta^{\mp}}_\lambda\cdot\O^{\beta^{\mp}}_\lambda\;
\O^{\beta^{\pm}}_\lambda\cdot\O^{\beta^{\pm}}_\lambda\ket{0^{+}},
\label{2.9}\end{eqnarray}
The double-charge-exchange sum rules (DSR) are:
\br
 S^{\beta\beta}_\lambda =S^{\beta\beta^{-}}_\lambda-S^{\beta\beta^{+}}_\lambda
=(2\lambda+1)^{-1}
\bra{0^{+}}[\O^{\beta^{+}}_\lambda\cdot\O^{\beta^{+}}_\lambda,
\O^{\beta^{-}}_\lambda\cdot\O^{\beta^{-}}_\lambda]\ket{0^{+}},
\label{2.10}\end{eqnarray}
which when evaluated give  \cite{Vog88,Mut92}:
\br
 S^{\beta\beta}_{F}\equiv S^{\beta\beta}_0 &=&2(N-Z)(N-Z-1),
\label{2.11}\end{eqnarray}
and
\br
 S^{\beta\beta}_{GT}\equiv  S^{\beta\beta}_1 &=&2(N-Z)\left(N-Z-1+2S^{\beta^+}_1\right)-\frac{2}{3}C,
\label{2.12}\end{eqnarray}
where $C$ is a relatively small quantity and is given by  \cite[(5)]{Mut92}.


\section{ Charge-exchange Quasiparticle Random Phase Approximation and Beyond}
The $\beta \beta$ decays occur in medium-mass nuclei that are often far from closed shells, and, as a
consequence, most of the recent attempts to evaluate ${\cal M}_{2\nu}$ and ${\cal M}_{0\nu}$
rely on the neutron-proton QRPA, because this model is much simpler computationally than the
shell model (SM). Note that the kind of correlations that these two methods include are not the same.
The QRPA deals with a large fraction of nucleons in a large single-particle space, but within a
modest configuration space. The shell model, by contrast, deals with a small fraction of nucleons
in a limited single-particle space, but allows them to correlate in arbitrary ways
within a large configuration space.

The charge-exchange QRPA has been first formulated, and applied to the allowed  $\beta$-decay and to
the collective GT resonance, by Halbleib and Sorensen in 1967 \cite{Hal67}. However, intensive
implementations of QRPA to $\beta\beta$-decay began only about 20 years later when Vogel
and Zirnbauer \cite{Vog86} discovered that the ground state correlations (GSC) play an essential
role in  suppressing the $2\nu\beta \beta$ rates. Soon afterwards, Civitarese, Faessler and
Tomoda \cite{Civ87} arrived to the same conclusion.  Almost simultaneously, Tomoda and  Faessler
\cite{Tom87}, and Engel, Vogel and Zirnbauer \cite{Eng88} revealed a similar though smaller
effect on the $0\nu\beta \beta$ decay.

When applied to the $\beta \beta$-decay the following two
steps are performed within the standard QRPA:
\bnu
\item Two charge-exchange  QRPA equations are solved for the
intermediate $1^+$ states; one  for the initial nucleus
$(N, Z)$ and one for the final nucleus $(N-2, Z+2)$. The first one,
\begin{eqnarray}
\left(\begin{array}{ll} A & B \\  B &
A\end{array}\right) \left(\begin{array}{l} X
 \\ Y \end{array}\right) =
\omega_\alpha \left(\begin{array}{l} ~X \\-Y \end{array}\right),
\label{3.1} \end{eqnarray}
is evaluated in the BCS vacuum
\be
\ket{{0}^+}=\prod_p(\up+\vp c^\dag_p c^\dag_{\bar p})
\prod_n(\un+\vn c^\dag_n c^\dag_{\bar n})\ket{},
\label{3.2}\ee
where $\ket{}$ stands for the particle vacuum. The Eq. \rf{3.1}
describes simultaneously four nuclei: $(N-1,Z-1),~(N+1,Z-1),
~(N-1,Z+1)$ and $(N+1,Z+1)$.
The
 matrix elements of the operators $\O^{\beta^\pm}_1$
are:
\begin{eqnarray}
\Bra{1^{+}_\alpha}\O^{\beta^-}_1\Ket{ 0^{+}}&=&\sum_{pn}
\left[{1}^0_+({pn};1) X_{{ pn};1^{+}_\alpha}
+{1}^0_-({pn;1}) Y_{pn;1^{+}_\alpha}\right],
\nn\\
\Bra{1^{+}_\alpha}\O^{\beta^+}_1\Ket{ 0^{+}}
&=&\sum_{pn}
\left[{1}^0_-({pn};1) X_{pn;1^{+}_\alpha}
+{1}^0_+({pn};1) Y_{pn;1^{+}_\alpha}\right],
\label{3.3}
\end{eqnarray}
were
\br
{\Lambda}^0_+({pn;\lambda})&=&\up \vn\Bra{p}{\rm O}_\lambda\Ket{n},
\nn\\
{\Lambda}^0_-({ pn;\lambda})&=&\vp \un \Bra{p}{\rm O}_\lambda\Ket{n},
\label{3.4} \er
are the unperturbed strengths.
The ISR reads
\begin{equation}
 S^{\beta}_{GT}=\frac{1}{3}
\sum_{\a}\left[|\Bra{1_\a^+}\O^{\beta^{-}}_1\Ket{0^+}|^2-
|\Bra{1_\a^+}\O^{\beta^{+}}_1\Ket{0^+}|^2\right]=N-Z.
\label{3.5}\ee

In the same way  the second QRPA  does not deal with the
$(N-1,Z+1)$ nucleus  only, but entangles as well the
isotopes
$(N-1,Z+3),~(N-3,Z+1)$ and $(N-3,Z+3)$.
The corresponding ISR is
\begin{equation}
\overline{ S}^{\beta}_{GT}=\frac{1}{3}
\sum_{\a}\left[|\Bra{\overline{1}_\a^+}\O^{\beta^{-}}_1\Ket{\overline{0}^+}|^2-
|\Bra{\overline{1}_\a^+}\O^{\beta^{+}}_1\Ket{\overline{0}^+}|^2\right]
=\overline{N}-\overline{Z}=N-Z-4,
\label{3.6}\ee
where the barred kets
($\ket{\overline{0}^+}$, $\ket{\overline{1}^+_\alpha}$)
indicate that the quasiparticles
are defined with respect to the  final nucleus.

\item
The equation \rf{2.1} is substituted by one of the following two ansatz:
\enu
\bit
\item  \underline{ Method M1} (proposed by Vogel and Zirnbauer \cite{Vog86,Eng88}):
\eit
\begin{eqnarray}
{\cal M}_{2{\nu}}& =& \frac{1}{2} \sum_\alpha
\left[\frac{\Bra{1^+_\alpha}\O^{\beta^+}_1\Ket{ 0^+}
\Bra{1^+_\alpha}\O^{\beta^-}_1\Ket{ 0^+}}{\w_{ 1^+_\alpha}}
+\frac{
\Bra{ \overline{1}^+_\alpha}\O^{\beta^+}_1\Ket{\overline{0}^+}
\Bra{\overline{1}^+_\alpha}
\O^{\beta^-}_1
\Ket{\overline{0}^+}}
{\w_{ \overline{1}^+_\alpha}}
\right],
\label{3.7}\end{eqnarray}
\bit
\item \underline { Method M2} (introduced  by Civitarese, Faessler and Tomoda \cite{Civ87,Tom87}):
\begin{eqnarray}
{\cal M}_{2{\nu}}& =&{2} \sum_{\alpha\alpha'}
\frac{\Bra{\overline{1}^+_{\alpha'}}\O^{\beta^+}_1\Ket{\overline{0}^+}
\ov{\overline{1}^+_{\alpha'}}{1^+_\alpha}
\Bra{1^+_\alpha}\O^{\beta^-}_1\Ket{ 0^{+}}}
{{\w}_{ 1^+_\alpha}+{\w}_{ \overline{1}^+_{\alpha'}}},
\label{3.8}\end{eqnarray}
where  the overlap in \rf{3.4} is evaluated as:
\be
\ov{\overline{1}^+_{\alpha'}}{1^+_\alpha}=
\sum_{pn}
\left[ X_{pn;1^{+}_\alpha}X_{pn;\overline{1}^{+}_{\alpha'}}-
Y_{pn;1^{+}_\alpha}Y_{pn;\overline{1}^{+}_{\alpha'}}\right].
\label{3.9}\ee
\eit

Note that the last two equations for ${\cal M}_{2{\nu}}$, \rf{3.7} and \rf{3.8}, cannot be derived
mathematically, and that they are just recipes which make the applications of the QRPA to the
$\beta \beta$-decay possible. Moreover, in many applications of the method M1, the energy
denominator $({\w}_{ 1^+_\alpha}+{\w}_{\overline{1}^+_{\alpha'}})/2$ in \rf{3.4} is simply taken
to be equal to $\omega_{ 1^+_\alpha}$. The GSC in $\Bra{1^+_\alpha}\O^{\beta^-}_1\Ket{ 0^{+}}$
and $\Bra{\overline{1}^+_{\alpha'}}\O^{\beta^+}_1\Ket{\overline{0}^+}$  are also different, matching,
respectively, transitions  $(N,Z)\betapgo (N+1,Z-1)$ and $(N-2,Z+2)\betago (N-3,Z+3)$.

When compared to the SM, the QRPA  presents the following drawbacks:
\bnu
\item [I)] There is ambiguity in treating the intermediate
states, and  further developments must be made to match the excited states of the odd-odd
nuclei based on  different  ground
states of the initial and final even-even nuclei, as  in \rf{3.7}
and \rf{3.8}.

\item [II)] The QRPA
$\beta \beta$-decay amplitudes are extremely  sensitive to the PN
interaction in the particle-particle ($pp$) channel, or more
precisely to
the ratio between the $S=1$ and $S=0$ forces
\cite{Vog86,Civ87,Tom87,Eng88,Hir90a,Hir90b,Krm92,Krm93,Krm93a,Krm94,Krm94a}.
\footnote{For a $\delta$ force this  ratio is
 $t=v_t^{pp}/v_s^{pp}$. Usually
is introduced in an  ad hoc parameter, denoted by $g^{pp}$.}
Even  worse, the model collapses as a whole
in the physical region of $t$ or $g^{pp}$.

\item [III)]
The QRPA is only capable to account for the $\beta\beta$-decay into the ground state
$\ket{0^+_1}$ of the final nucleus, while to deal with  final excited states it is
necessary to recur to another nuclear model. Usually,  one solves an extra
charge-conserving QRPA equation of motion for the $2^+$ -excitations on the final
BCS vacuum, and one assumes the final states $\ket{2^+_1}$ and $\ket{0^+_2}$ to be
the one-phonon quadrupole vibration, and a member of the two-phonon quadrupole vibrational
triplet, respectively  \cite{Gri92,Rad91,Rad96,Rad00,Sim01}.

\item [IV)]
In the QRPA we cannot evaluate   the energy distributions of the double
charge-exchange transition strengths, given by \rf{2.7},
\enu

There have been many tries to avoid the collapse
in  the QRPA, so its sensitivity to the phenomenological parameters
would become more realistic.
A first step in this direction has been done in Ref. \cite{Hir90a},
shortly after the finding of Vogel and Zirnbauer \cite{Vog86},
by working out  a  two-vacua QRPA (TVQRPA) specially
tailored  for the $\beta \beta$-decay. We called it so because one solves
the  QRPA equation of motion on  the quasiparticle vacuum
\be
\ket{\tilde{0}^+}=\prod_p(\up+\vbp c^\dag_pc^\dag_{\bar p})
\prod_n(\ubn+\vn c^\dag_nc^\dag_{\bar n})\ket{},
\label{3.10}\ee
which involves the vacua of both the initial ($u,v$) and final
(${\bar u},{\bar v}$) nuclei.

The forward $(\widetilde{A})$ and the backward $(\widetilde{B})$ going matrix elements are given by
\begin{eqnarray}
\widetilde{A}(pn,p'n';J)& =&(\widetilde{\e_p}+\widetilde{\e_n})\delta_{pn,p'n'}
\nn\\
&+&\sqrt{\rop \ron\ropp \ronp}[(\up\vn\upp\vnp
+\vbp \ubn\vbpp \ubnp) {\rm F}(pn,p'n';J)
\nn\\
&+& (\up\ubn\upp\ubnp+\vbp\vn\vbnp\ubnp){\rm G}(pn,p'n';J)],
\label{3.11} \end{eqnarray}
and
\br
\widetilde{B}(pn,p'n';J)&=&\sqrt{\rop \ron\ropp \ronp}
\left[(\vbp \ubn \upp \vnp + \up \vn \vbpp \ubnp ) {\rm F}(pn,p'n';J)\right.
\nn\\
&-& \left.(\up \ubn \vbpp \vnp + \vbp \vn \upp \vbnp ) {\rm G}(pn,p'n';J) \right].
\label{3.12} \end{eqnarray}
The quasiparticle energies are:
\be
\widetilde{\e_p}=\frac{\widetilde{\Delta}_p}{2\up\vbp\rop},\hspace{1cm}
\widetilde{\e_n}=\frac{\widetilde{\Delta}_n}{2\un\vbn\ron},
\label{3.13} \ee
with the pairing gaps given by
\br
\widetilde{\Delta}_p&=&-\frac{1}{2}\sum_{p'}
\sqrt{\frac{2j_{p'}+1}{2j_p+1}}\upp\vbpp\ropp G(pp,p'p';0),\hspace{1cm}
\nn\\
\widetilde{\Delta}_n&=&-\frac{1}{2}\sum_{n'}
\sqrt{\frac{2j_{n'}+1}{2j_n+1}}\unp\vbnp\ronp G(nn,n'n';0).
\label{3.14} \er
The factors  $\rop$  and  $\ron$ are defined as
$\rop^{-1}=\up^{2}+\vbp^{2}$, $\ron^{-1}=\ubn^{2}+\vn^{2}$, while the remaining notations have
standard meanings \cite{Hir90,Krm92}.

\begin{figure}[h]
\begin{center}
   \leavevmode
   \epsfxsize = 12cm
     \epsfysize = 5cm
    \epsffile{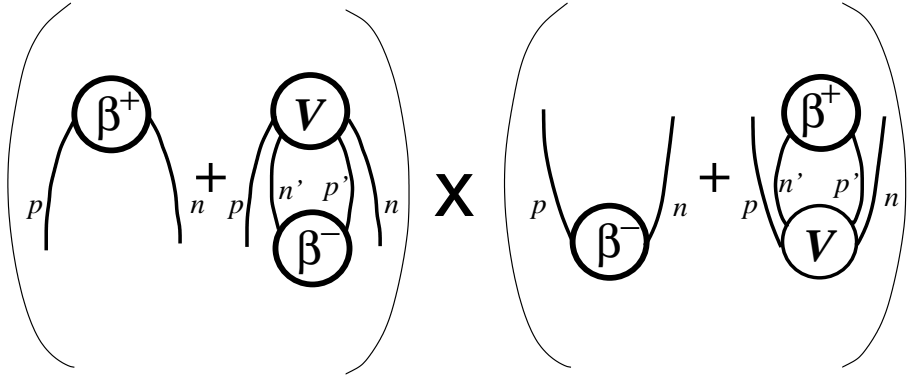}
\end{center}
\caption{\footnotesize
Graphical representation of the numerator in \rf{2.1} within the
QRPA.  The first
and second terms match, respectively, to
$\Bra{0^+_1}\O^{\beta^-}_\lambda\Ket{\lambda^+_\alpha}$ and
$\Bra{\lambda^+_\alpha}\O^{\beta^-}_\lambda\Ket{0^+}$.
The residual interaction  brings about the ground-state correlations in the vertex $V$.
}
\label{fig2}\end{figure}

The ${\cal M}_{2{\nu}}$ amplitude reads
\begin{eqnarray}
{\cal M}_{2\nu}&=&\sum_\a
\frac{
\Bra{\widetilde{1}^+_\alpha}\O^{\beta^+}_1\Ket{\widetilde{0}^+}
\Bra{\widetilde{1}^+_\alpha}\O^{\beta^-}_1\Ket{\widetilde{ 0}^+}}
{\widetilde{\w}_{1^+_\alpha}}
\label{3.15}\er
where
\begin{eqnarray}
\Bra{\widetilde{1}^+_\alpha}\O^{\beta^-}_1\Ket{\widetilde{ 0}^+}&=&\sum_{pn}
\left[\widetilde{\Lambda}^0_+({pn};1)\widetilde{ X}_{{ pn};1^{+}_\alpha}
+\widetilde{\Lambda}^0_-({pn;1})\widetilde{ Y}_{pn;1^{+}_\alpha}\right],
\nn\\
\Bra{\widetilde{1}^+_\alpha}\O^{\beta^+}_1\Ket{\widetilde{0}^+}
&=&\sum_{pn}
\left[\widetilde{\Lambda}^0_-({pn};1)\widetilde{ X}_{{ pn};1^{+}_\alpha}
+\widetilde{\Lambda}^0_+({pn;1})\widetilde{ Y}_{pn;1^{+}_\alpha}\right],
\label{3.16}
\end{eqnarray}
are, respectively, the perturbed $\beta^-$ and $\beta^+$
matrix elements, and
\br
\widetilde{\Lambda}^0_+({pn;\lambda})&=&\sqrt{\rop \ron}\up
\vn\Bra{p}{\rm O}_\lambda\Ket{n},
\nn\\
\widetilde{\Lambda}^0_-({ pn;\lambda})&=& \sqrt{\rop \ron}\vbp \ubn\Bra{p}{\rm
O}_\lambda\Ket{n},
\label{3.17} \er
are the unperturbed ones.
The ISR is now
\begin{equation}
\widetilde{ S}^{\beta}_{GT}=\frac{1}{3}
\sum_{\a}\left[|\Bra{\widetilde{1}_\a^+}\O^{\beta^{-}}_1\Ket{\widetilde{0}^+}|^2-
|\Bra{\widetilde{1}_\a^+}\O^{\beta^{+}}_1\Ket{\widetilde{0}^+}|^2\right]
\cong N-Z-2.
\label{3.18}\ee
Now and then the gap equations are solved for the intermediate nucleus
 in which  case the ISR gives exactly $N-Z-2$ \cite{Krm97,Bar97,Bar99}.

In Figure 2 we exhibit
the graphical representation of different terms within
the numerator in  \rf{3.13}.
The TVQRPA involves only the virtual $(N - 1,Z + 1)$-states, being the
backward-going contributions for the $\beta^-$ transitions the forward-going
contributions for the $\beta^+$ transitions, and vice versa.
\newpage
Based on a numerical comparison,  we have found that   the behavior of $2\nu$ transition amplitude
in the TVQRPA, with regard to the $pp$
 coupling, is   quite similar to that  previously noticed in the QRPA \cite{Hir90}.
Since no uncertainties are present when  intermediate states are treated in the TVQRPA,
 this result can, in some ways, be interpreted as a justification for  the averaging procedure
 performed in Eqs. \rf{3.7} and \rf{3.8}.

It could also be worth noting that in the TVQRPA we can express the amplitude
${\cal M}_{2\nu}$  in the form:
\begin{eqnarray}
{\cal M}_{2{\nu}}=\frac{1}{2}\left(\widetilde{\Lambda}^0_+,\widetilde{\Lambda}^0_-\right)\left(\begin{array}{cc}
 \widetilde{A} &  \widetilde{B} \\  - \widetilde{B} &
- \widetilde{A}\end{array}\right)^{-1} \left(\begin{array}{c}\widetilde{\Lambda}^0_+
 \\ -\widetilde{\Lambda}^0_- \end{array}\right),
\label{3.19} \end{eqnarray}
and can therefore  calculate the transition probability
 without first solving  the QRPA equation.

We found the last  equation  to be  especially useful for discussing
the $\beta
 \beta$-decay rates within the single-mode-model (SMM)
\cite{Krm92, Krm93, Krm94a},
which    deals with
only one intermediate state for
each $J^{\pi}$ and is the simplest version of the QRPA.
Within the  SMM one can express   the moment ${\cal M}_{2\nu}$
 \`{a} la Alaga \cite{Ala71},
\begin{eqnarray}
{\cal M}
_{2\nu}&=& g^{eff}_{2\nu}{\cal M}_{2\nu}^0,
\label{3.20}\end{eqnarray}
\ie  as the unperturbed BCS matrix element
\be
 {\cal M}_{2\nu}^0=\up\vn\ubn\vbp\rop \ron\frac{ \Bra{p}O_1\Ket{n}^2}{\omega^{0}},
\label{3.21}\ee
 multiplied by an  effective charge:
\begin{eqnarray}
g^{eff}_{2\nu}&=& \left(\frac{\omega^{0}}
{{\omega}_{1^{+}}}
\right)^{2}\, \left[1+\,\frac{{\it G}(pn,pn;1^+)} {{\omega}^{0}}\right],
\label{3.22}
\end{eqnarray}
where
\begin{eqnarray}
\omega^{0}=-\frac{1}{4}\left[{\it G}(pp,pp;0^+) +{\it G}(nn,nn;0^+)\right],
\label{3.23}
\end{eqnarray}
is the unperturbed pairing energy between protons and neutrons.
$g^{eff}_{2\nu}$ comes from the QRPA correlations, or more precisely from
the interference between the forward and backward going contributions,
which add coherently in the $pp$ channel and totally out of phase in
the $ph$ channel.

Still more, as the QRPA energy  ${\omega}_{1^{+}}$ is  in essence a linear
function of ${\it G}(1^{+})/\omega^{0}$,
we can  state that, because of the GSC, the  effective QRPA charge
$g^{eff}_{2\nu}$ is mainly a bilinear function of
${\it G}(1^{+})/\omega^{0}$, or equivalently of $t$
\cite{Krm93,Krm94},  \ie
\be
g^{eff}_{2\nu}\sim
\frac{1-t/t_0}{1-t/t_1},
\label{3.24}\ee
That is,  ${\cal M}_{2\nu}$  passes   through zero at $t=t_0$ where
${\it G}(1^{+})=-{\omega}^0$, and  has a
pole at $t=t_0$ where ${\omega}_{1^{+}}=0$.

The fact that in many situations the SMM reproduces quite well the complete QRPA calculations
made us suspect  that the behavior of ${\cal M}_{2\nu}$ with regard to $t$ could always  follow a
$(1,1)$-Pad\'{e} approximant of the form \rf{3.21}.
Thus, we have  suggested that, independently of the nuclear Hamiltonian and/or the
configuration space employed in the QRPA calculation, the $2\nu$
amplitude should unavoidably behave as \cite{Krm93}
\be
{\cal M}_{2\nu}= {\cal M}_{2\nu}(t=0)
\frac{1-t/t_0}{1-t/t_1}.
\label{3.25}
\ee
At variance with the bare BCS
value ${\cal M}_{2\nu}^0$, given by \rf{3.20},
the matrix element  ${\cal M}_{2\nu}(t=0)$ also contains the $ph$-like
 correlations.
We have tested the relation \rf{3.24} only for  a zero-range force
\cite{Krm93}, but as far as I know there is no QRPA calculation
in the literature that could be in conflict with this result
 \footnote[1] {When the renormalization coupling constant  $g^{pp}$ is used \cite{Civ87} a similar
 expression to the Eq. (\ref{3.22}) is valid (with $g^{pp}$'s for $t$'s).}.

One can also guess  that the breakdown in the QRPA  comes from the  violation  of  the particle-number
symmetry, caused by the BCS  approximation. Therefore with the hope of getting out of this
inconvenience we have worked out a particle number projected QRPA (PQRPA) for the  charge-exchange
excitations starting  from the time-dependent variational principle,  \cite{Krm93a}. However, after
performing numerical calculations for the $2\nu\beta\beta$-decay in $^{76}Ge$, we found that
in this model  the ${\cal M}_{2\nu}$ amplitude continues to behave
roughly as in the plain  QRPA. Said in another way,  the number projection procedure is unable to avoid the collapse.

The  variation  of the QRPA  that has received
major attention lately is the so-called renormalized QRPA (RQRPA)
\cite{Toi95,Krm96,Sch96,Krm97,Toi97,Sim97,Mut97,Eng97}.
The new ingredient brought up by this model is the effect of the GSC
 in the QRPA equation of motion itself.  The important outcome of this is that
 the QRPA collapse  does not develop anymore in the physical region
of the {\it pp}-strength parameter.
Yet,  this new procedure  to incorporate the GSC tones down  only slightly
the strong dependence of the
$2\nu\beta\beta$ transition amplitude on this parameter.

On the other hand  we soon found  \cite{Krm96} that the price one has to pay  in the RQRPA to  avoid
the collapse was the non-conservation of the ISR, given by  the Eq. \rf{2.6}. This violation is about
$20-30\%$ and  we cannot  get away from it. It comes from the fact that the scattering part of the
GT operator,  when acting on the RQRPA ground state, creates states that are not contained in the
model space. These terms have recently been considered in the framework of the
``Fully Renormalized QRPA" (FR-QRPA) \cite{Rod02}, whereby the ISR was successfully restored.
Yet, within the FR-QRPA  the  $2\nu\beta\beta$ amplitude behaves similarly as in the ordinary QRPA.
Namely, in this model ${\cal M}_{2\nu}$ passes through zero and develops a
pole for values of the {\it pp}-strength parameter which are only slightly  higher than those in the QRPA model.

Let us also remember that only  the self-consistent QRPA (SCQRPA) theory  \cite{Duk90,Duk98,Krm98,Pas98}
 incorporates fully the GSC, leading simultaneously  to a coupling of the single-particle field to the QRPA
 excitations. We performed \cite{Krm98} a detailed comparison of the properties of the QRPA, RQRPA and SCQRPA
  equations in the $O(5)$ model for the F excitations, inferring that:
 i) before the QRPA collapses all three approaches reproduce well correct results,
 ii) near the transition point only the SCQRPA values are close to the exact ones, and iii) beyond that point
 both the SCQRPA and RQRPA yield values different from the exact ones, but the former are somewhat better.
  One can suspect that in realistic cases this condition prevails and, to some extent, even for the GT transitions.
 Such a  possibility has not been explored so far.
It should also be mentioned  that what some authors \cite{Bob99} refer to as self-consistent QRPA is just
the RQRPA with the introduction of some minor changes, given by Eqs. (34), (40) and (45) in Ref. \cite{Krm97}.
\footnote{See also  comments with regard to this in Ref. \cite{Mar00}.}

We arrive therefore to the conclusion that not one of the amendments of the
QRPA, proposed so far to rescue this nuclear model, was  able to
change qualitatively the
behavior of the amplitude ${\cal M}_{2\nu}$, given by \rf{3.22},   unless
we agree to tolerate  the violation of the ISR,  which
could be extremely dangerous as we have no control on
how it affects the definite value of ${\cal M}_{2\nu}$.
More, neither the RQRPA nor the SCQRPA is able to
evade the other three unfavorable  QRPA outcomes that we have
 pointed out at the beginning of this section.
Thus,   within this
QRPA scenario, instead of introducing further improvements and
variations
into the QRPA equation of motion, it would perhaps  be a good idea
 ``to shuffle the
cards and deal again'', \ie to try to work out a different quenching
mechanism for the NME's.  In the next two sections we  explore that
possibility.

\section{ Quasiparticle Tamm-Dancoff  Approximation (QTDA)}

Here we  sketch a simple nuclear model for evaluating the $\beta\beta$
decay rates, based
on  the well-known QTDA
\cite{Pal66,Ram69}.  It will become clear immediately that the
main difference in comparison  to the QRPA comes from how one
describes the final $(N-2, Z+2)$ nucleus.

Same as   in the QRPA, we conveniently express the total Hamiltonian  as
\be
H=H_p+H_n+H_{pn}+H_{pp}+H_{nn}\equiv H_0+H_{res},
\label{4.1}\ee
where $H_p$ and  $H_n$  are the effective proton and neutron single-quasiparticle Hamiltonians
(with eigenvalues ${\e_p}$ and ${\e_n}$), while  $H_{pn}$, $H_{pp}$, and $H_{nn}$ are the
matching effective two-quasiparticle interaction Hamiltonians among the valence quasiparticles.

We assume both i) that the initial state is the $BCS$ vacuum in the
$(N, Z)$  nucleus, and ii) that  the intermediate and final nuclear
states involved in the
 $\beta\beta$-decay are, respectively, two and four
 quasiparticle excitations  on this vacuum. That is:
\bit
\item{initial state: $\ket{0^+}=\ket{BCS}$},
\item{intermediate states}:
\be
\ket{\lambda^{+}_\alpha}=\sum_{pn} X_{pn;\lambda^{+}_\alpha}\ket{pn;\lambda^+},
\label{4.2}\ee
with
\be
\ket{pn;\lambda^+}=[a^\dag_pa^\dag_n]_{\lambda^+}\ket{BCS},
\label{4.3}\ee
\item{final states}:
\be
\ket{0^{+}_f}=\sum_{p_1p_2n_1n_2J} Y_{{ p_1p_2n_1n_2
J};0^{+}_f}\ket{p_1p_2,n_1n_2;J},
\label{4.4}\ee
with
\be
\ket{p_1p_2,n_1n_2;J}=N(p_1p_2)N(n_1n_2)\{
[a^\dag_{p_1}a^\dag_{p_2}]_{J}
[a^\dag_{n_1}a^\dag_{n_2}]_{J}\}_{0}\ket{BCS},
\label{4.5}\ee
and
\be
N(ab)=(1+\delta_{ab})^{-1/2}.
\label{4.6}\ee
Here   $a^{\dagger}$ ($a$) is the quasiparticle creation
(annihilation) operator relative  to the BCS vacuo.
\eit
\begin{figure}[h]
\begin{center}
   \leavevmode
   \epsfxsize = 5cm
     \epsfysize =6cm
    \epsffile{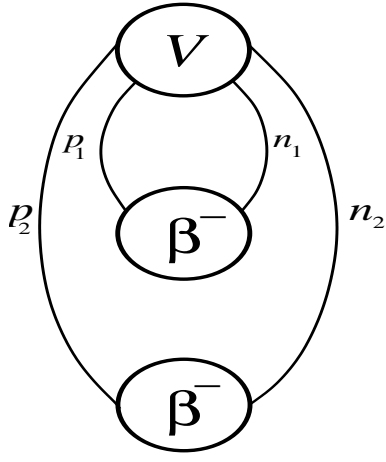}
\end{center}
\caption{\footnotesize
Graphical representation of ${\cal M}_{2\nu}$ in the QTDA. The first and the second vertex
match with matrix elements \rf{4.7} and \rf{4.8}, respectively, while the third vertex
represents the residual interaction in the final state.
}
\label{fig3}\end{figure}
One can read the matrix elements of $H_{pn}$ between the intermediate states $\ket{pn;\lambda^+}$ and
the  final states $\ket{p_1p_2,n_1n_2;J}$, respectively  from   \cite[(4.9) and (4.5)]{Ram69}, and
those of $H_{pp}$ and $H_{nn}$ between the same final states  from  \cite[(2.11)]{Pal66}.
We show explicit results only for the one-body  matrix elements that appear in \rf{2.1}. They are:
\begin{eqnarray}
\Bra{\lambda^{+}_\alpha}\O^{\beta^-}_\lambda\Ket{ 0^{+}}&=&\sum_{pn}{\Lambda}^0_+({pn;\lambda})
 X_{pn;\lambda^{+}_\alpha},
\label{4.7}\er
and
\br
\Bra{ 0^{+}_f}\O^{\beta^-}_\lambda\Ket{\lambda^{+}_\alpha}&=&-\sum_{pn}X_{pn;\lambda^{+}_\alpha}
\sum_{p_1p_2n_1n_2J} Y_{{\rm p_1p_2n_1n_2J};0^{+}_f}
N(p_1p_2)N(n_1n_2){\bar P}(p_1p_2J){\bar P}(n_1n_2J)
\nn\\
&\x&
\sqrt{2J+1}(-)^{n_1+p_2+J+\lambda}\sixj{p_1}{n_1}{\lambda}{p}{n}{J}
{\Lambda}^0_+({p_1n_1;\lambda})\delta_{p_2p}\delta_{n_2n},
\label{4.8}
\end{eqnarray}
where
\br
{\bar P}(p_1p_2J)=1-(-)^{p_1+p_2+J}P(p_1\leftrightarrow p_2),
\label{4.9}\er
is the well known permutation operator.
The energies in the  denominator \rf{2.1} are
\begin{eqnarray}
E_{\lambda^+_\alpha }&=&E_0+\w_{\lambda^+_\alpha}+\lambda_p-\lambda_n,
\nn\\
E_{0^+_f}&=&E_0+\w_{0^{+}_f}+2\lambda_p-2\lambda_n,
\label{4.10}\end{eqnarray}
where $\w_{\lambda^{+}_\alpha }$ and $\w_{0^{+}_f}$ are the eigenvalues of the Hamiltonian \rf{4.1}
for intermediate states  $\ket{\lambda^+_\alpha }$ and final states $\ket{0^+_f}$, respectively,
and $\lambda_p$  and $\lambda_n$ are the chemical potentials. Therefore
\br
{\cal D}_{\lambda^{+}_\alpha,f } &=&\w_{\lambda^{+}_\alpha
}-\frac{\w_{0^{+}_f}}{2}.
\label{4.11}\end{eqnarray}

The QTDA has the correct particle-hole (shell model) limit:
$v_p\go 0, v_n\go 1$, and therefore one  can  straightforwardly apply this model
to single- and double-closed shell nuclei. In the next section  we discuss an example.

\section{ $2\nu\beta\beta$-decay $^{48}Ca \rightarrow\, ^{48}Ti$}

There are two recent experimental results for the
$2\nu\beta\beta$-decay half-life to the $0^{+}_1$ state that nicely
agree with each other. They are:
\br
 \mbox{Ref. \cite{Bal96}}:\hspace{1cm}T_{2\nu}&=&\left
(4.3^{+2.4}_{-1.1}[\mbox{stat}]\pm 1.4[\mbox{syst}]\right)\x 10^{19} \mbox{ yr},
\nn\\
 \mbox{Ref. \cite{Bru00}}:\hspace{1cm}T_{2\nu}&=&\left
(4.2^{+3.3}_{-1.3}\right)\x 10^{19} \mbox{ yr},
\label{5.1}\er
which, from \rf{1.5} and the kinematical factor \cite{Krm94},
\be
{\cal G}_{2\nu}=42.3 \x 10^{-19} \left[\mbox{ yr(MeV)}^2\right]^{-1},
\label{5.2}\ee
yield:
\br
 \mbox{Ref. \cite{Bal96}}:\hspace{1cm}|{\cal M}_{2\nu}|&=&
\left(0.074^{+0.040}_{-0.020}\right) [\mbox{MeV}]^{-1},
\nn\\
 \mbox{Ref. \cite{Bru00}}:\hspace{1cm}|{\cal M}_{2\nu}|&=&
\left(0.075^{+0.015}_{-0.019}\right) [\mbox{MeV}]^{-1}.
\label{5.3}\er
One should keep in mind  that the experimental values for $|{\cal M}_{2\nu}|$ in  \rf{5.3}
depend, through the value of ${\cal G}_{2\nu}$, on the value used for the effective
axial-vector coupling constant $g_{A}$. In the present work we use $g_{A}=1$. Clearly for the bare
value, $g_{A}=1.26$, the phenomenological NME's decrease by factor $(1.26)^2$.

There is also a very interesting high-resolution charge-exchange reaction experiment \cite{Rak04},
  where  ${\cal M}_{2\nu}$ for the ground state in $^{48}$Ti, was built from energy spectra of
  the $^{48}$Ca (p, n)$^{48}$Sc and $^{48}$Ti(d,$^2$He)$^{48}$Sc reactions, by converting the
  (p,n) and (d,$^2$He) cross sections into moments $\Bra{0^+_1}\O^{\beta^-}_1\Ket{1^+_\alpha}$
  and $\Bra{1^+_\alpha}\O^{\beta^-}_1\Ket{0^+}$ which contribute in \rf{2.1}. In performing the
  summation over $\alpha$ five experimentally observed states below $5$ MeV have been considered,
  under the assumption that all matrix elements are positive. In this way Rakers \etal  \cite{Rak04} get:
\br
 \mbox{Ref. \cite{Rak04}}:\hspace{1cm}|{\cal M}_{2\nu}|_{E\le 5 \,{\rm  MeV}} &=&
\left(0.0740\pm 0.0150\right) [\mbox{MeV}]^{-1}.
\label{5.4}\er
One should note   that the previous result does not depend on the value used for $g_{A}$, and that although
\rf{5.3} and \rf{5.4} agree numerically they are physically different.

\begin{figure}[h]
\begin{center}
   \leavevmode
   \epsfxsize = 11cm
     \epsfysize = 9cm
    \epsffile{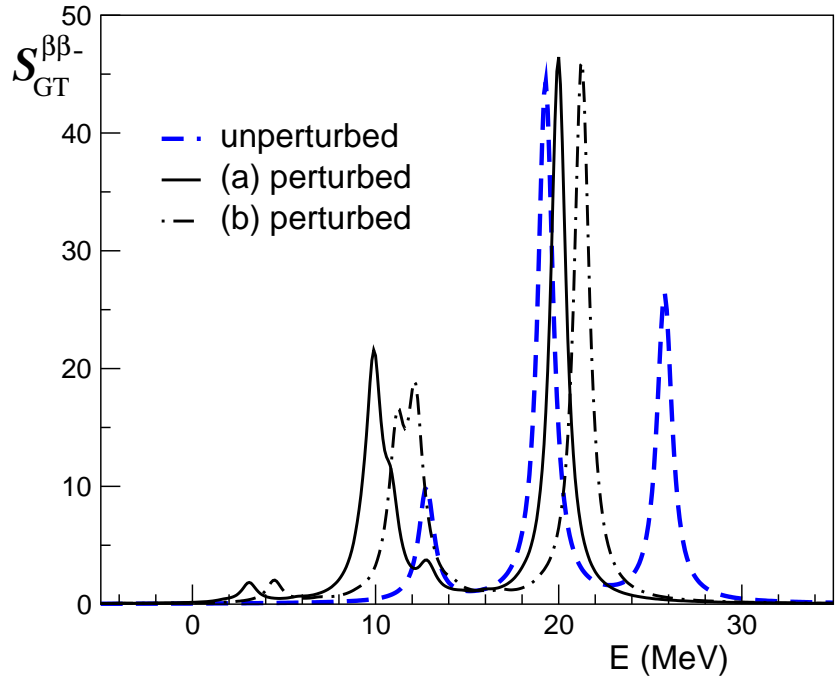}
\end{center}
\caption{\footnotesize
Graphical representation of energy dependence of the  double  GT strengths $ S^{\beta\beta}_{GT}$ ,
measured from the $^{48}$Ca  ground state. The coupling constant values in the perturbed calculations
were  $v_s=35$, $v_t=65$ for case (a) , and   $v_s=40$, $v_t=60$ for case (b).}
\label{fig4}\end{figure}

Previously, jointly with Dubravko, we had already studied the $\beta\beta$ decay  in  $^{48}$Ca, but only
 for the single particle state $1f_{7/2}$ \cite{Bar99}. Here we consider the complete $pf$
 particle-hole space. Of course,  this  configuration space is still strongly limited when compared to
 those of the SM within the $pf$  single-particle space \cite{Zha90, Cau94, Pov95}. Thus, obviously, a SM
 yields a more realistic  description of the $\beta\beta$-decay  \cite{Bal96,Bru00} and of the
 related charge-exchange
reactions  \cite{Rak04} than the present model.
Our only purpose here is to explain   the  way  the QTDA quenching-mechanism works.
Choosing suitable effective single-particle energies (SPE) is, as always, a delicate
issue in nuclear structure calculations. We will take them to be:
\br
\e_{j_n}&=&\e^0_{j_n}+\mu_{j_n},
\nn\\
\e_{j_p}&=&\e^0_{j_n}+\mu_{j_p}+\Delta_C,
\label{5.5}\er
where $\e^0_{j_n}$ are experimental SPE for  $^{40}Ca$,  extracted from Fig. 2 in Ref. \cite{Bro98},
namely: $ \e^0_{f_{7/2}}=0$,  $ \e^0_{f_{5/2}}=6.5$, $ \e^0_{p_{3/2}}=2.1$, and $ \e^0_{p_{1/2}}=4.1$,
in units of MeV. The self-energies $\mu_{j_n}$ and $\mu_{j_p}$  account for $8$ neutrons in the $f_{7/2}$
shell, and are defined in Ref. \cite{Krm97}. $\Delta_C$ is the Coulomb displacement energy in $^{48}Ca$.
With SPE chosen in this way  the energy of the isobaric analog state (IAS) is always equal to  $\Delta_C$,
as it should be.

We will use the $\delta$-force
\be
V=-4\pi(v_sP_s+v_tP_t)\delta(r)
\label{5.6}\ee
both for evaluating the self-energies in \rf{5.5} and for calculating the residual interaction.
Besides, for the purpose of simplifying the discussion,  we will assume the coupling strengths
$v_s$ and $v_t$ to be equal for identical and for different particles.

Let us first comment the unperturbed results, \ie when $v_s=v_t=0$. In this case all  double F
$\beta\beta$-strength, $ S^{\beta\beta}_F\equiv S^{\beta\beta^-}_F=124$, concentrates in the
lowest lying  degenerate states $\ket{f_{7/2}^2, f_{7/2}^2; J=0,2,4,6 }$, in parts of
$S^{\beta\beta^-}_{F}(J)=4(2J+1)$. These are the  double IAS's (DIAS's) and are at energy
$2\Delta_C=12.78$ MeV,  measured from  the  $^{48}$Ca  ground-state,  which was taken to be
  $E_0=0$.
Contrarily, we found in these  states only $12\%$  of the total double  GT  intensity,
  $ S^{\beta\beta}_{GT}\equiv S^{\beta\beta^-}_{GT}=125.71$,
  being equal  to: $2.20$,  $7.22$, $2.64$ and $3.18$ for $J=0,2,4$ and $6$,  respectively.
  The major part of the double GT strength concentrates at levels $\ket{f_{7/2}^2,f_{7/2}
   f_{5/2}; J=2,4,6 }$ ($55\%$) and $\ket{f_{7/2}^2,f_{5/2}^2; J=0,2,4 }$ ($33\%$), which, as seen
   from  Figure 4 lie, respectively,  at energies  $2\Delta_C+\e^0_{f_{5/2}}$, and
   $2\Delta_C+2\e^0_{f_{5/2}}$. For the lowest final $0^+$ state the unperturbed denominators
   \rf{4.10}  are all null, which makes  both NME's, $ {\cal M}_{2{\nu}}^{F}$ and
   ${\cal M}_{2\nu}^{GT}$, to become $\infty$.   The scene   changes radically
   when the residual interaction is switched on. First, as we show in Figure 5, there is a
   great variety of physically sound values for $v_s$ and $v_t$ that allow the model  to account
   for the phenomenological NME's \rf{5.3} and \rf{5.4}. In another words, same as the QRPA, the
    QTDA is capable of restoring the Wigner SU(4) symmetry, quenching in   this way   the NME's.
    This is precisely  the mechanism  we have been searching for.

\begin{figure}[t]
\begin{center}
   \leavevmode
   \epsfxsize = 10cm
     \epsfysize = 9cm
    \epsffile{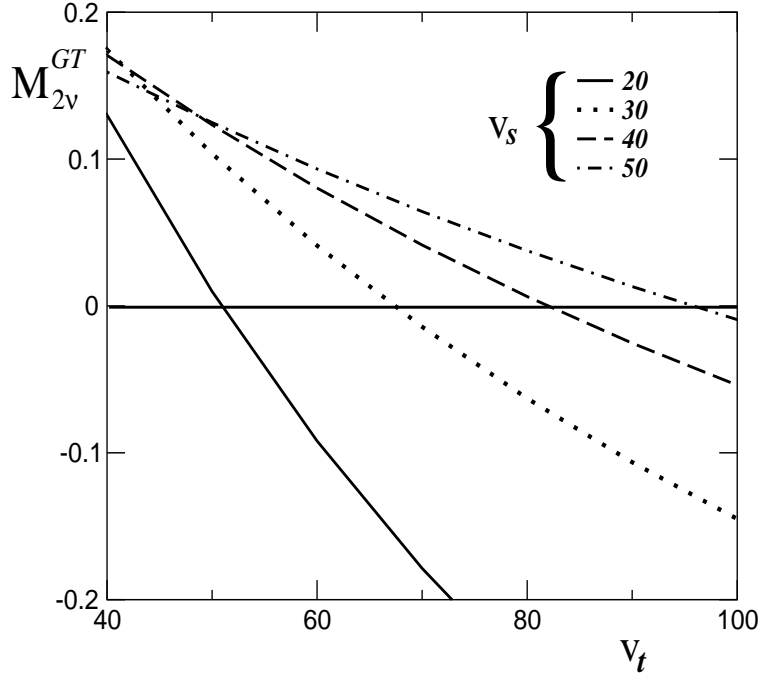}
\end{center}
\caption{\footnotesize
Graphical representation of the $\beta\beta$ amplitudes
 ${\cal M}^{GT}_{2\nu}$ in the QTDA as a
function of coupling constants $v_s$ and $v_t$.}
\label{fig5}\end{figure}

\begin{figure}[t]
\begin{center}
   \leavevmode
   \epsfxsize = 8cm
     \epsfysize = 8cm
    \epsffile{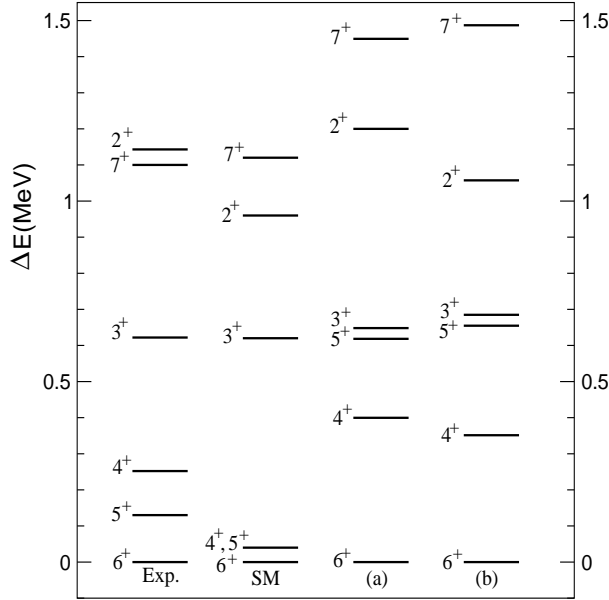}

\end{center}
\caption{\footnotesize
Comparison between the experimental and theoretical energy levels of
$^{48}$Sc for the cases (a) $v_s=35$, $v_t=65$, and (b) $v_s=40$, $v_t=60$.
We also  display  the shell-model results from Ref. \cite{Cau94}.
}
\label{fig6}\end{figure}

 Just for the sake of illustration
 we  use here two  sets of  coupling constants:
 (a) $v_s=35$, $v_t=65$, and (b) $v_s=40$, $v_t=60$, which
reproduce reasonably well the energy levels of $^{48}$Sc, shown in Figure 6,
and are close to  values   used    in our previous works
\cite{Krm94,Krm97,Hir90,Bar99} within the
PN ph channel ($v_s^{ph}=27$ and $v_t^{ph}=64$). Note first that, as
$\bra{f_{7/2},f_{5/2};1^+}H_{pn}\ket{f_{7/2},f_{7/2};1^+}\sim (v_s+v_t
 )$, the wave function for the GTR is the same in both cases:
\br
\ket{\rm GTR}&=& 0.972 \ket{f_{7/2},f_{5/2};1^+}
-0.237 \ket{f_{7/2},f_{7/2};1^+}.
\label{5.7}\er
The residual interaction acts differently  on double F and on GT transition strengths.
In the first case, all  intensity remains concentrated at the energy $2\Delta_C$,
independently of the values of $v_s$ and $v_t$, but now there is only  one DIAS.
Its wave function  is a coherent superposition of the states $\ket{f_{7/2}^2,f_{7/2}^2; J=0,2,4,6 }$, \ie
\be
\ket{{\rm DIAS}}=\sum_J\sqrt{\frac{2J+1}{\sum_J(2J+1)}}
\ket{f_{7/2}^2,f_{7/2}^2; J}.
\label{5.8}\ee
In the second case  the double GT resonance (DGTR),  placed at an
energy that is only slightly lower  than
 $2\Delta_C+\e^0_{f_{5/2}}$, carries  a large part of  the strength.
That is: $S^{\beta\beta^-}_{GT}$ (DGTR)$ =72.87$ in  case (a) and $=72.61$
in case (b). The DGTR   wave function is:
\br
\ket{\rm DGTR}&=&
\left.\begin{array}{ll} 0.405 \\0.371 \end{array}\right\}
\ket{f_{7/2}^2,f_{5/2}^2; 0}
+\left.\begin{array}{ll} 0.683 \\0.683 \end{array}\right\}
 \ket{f_{7/2}^2,f_{5/2}^2; 2}
+\left.\begin{array}{ll} 0.505 \\0.533 \end{array}\right\}
\ket{f_{7/2}^2,f_{5/2}^2; 4}+\cdots .
\nn\\
\label{5.9}\er
The remaining GR strength mainly comes from the
states $\ket{f_{7/2}^2,f_{7/2} f_{5/2}; J=2,4,6}$,
and in both cases  concentrates almost fully at the energy of $\sim 10-11$ MeV.
The unperturbed and perturbed double GT strengths are confronted in Figure 4.
The calculated wave function for  the ground state in $^{48}Ti$
is
\br
\ket{0^+_1}&=&
\left.\begin{array}{ll} 0.922\\0.922 \end{array}\right\}
 \ket{f^2_{7/2},f^2_{7/2}; 0}
-\left.\begin{array}{ll} 0.250 \\0.226 \end{array}\right\}
\ket{f^2_{7/2},f^2_{7/2}; 2}
-\left.\begin{array}{ll} 0.095 \\0.093 \end{array}\right\}
\ket{f^2_{7/2},f^2_{7/2}; 4}
\nn\\
&+&
\left.\begin{array}{ll} 0.189 \\0.211\end{array}\right\}
\ket{f^2_{7/2},f^2_{5/2}; 0}
+\left.\begin{array}{ll} 0.119\\0.138 \end{array}\right\}
\ket{f^2_{7/2},p^2_{3/2}; 0}+\cdots.
\label{5.10}\er
It   appears at the energy $E_{0^{+}_1}=-4.41$ MeV in case (a) and at
$-3.95$ MeV in case (b), which compares favorably with the
measured value $Q_{\beta\beta^-}=4.27$ MeV \cite{Bal96,Bru00,Rak04}.

\begin{table}[h]
\begin{center}
\caption
{Decomposition  of the total GT matrix elements. The denominators
${{\cal D}_{1 ^+_\alpha,1}}$ are given in units of MeV, and
the moments ${\cal M}_{2\nu}^{GT}$ in units of (MeV)$^{-1}$.    }
\label{table1}
\bigskip
\begin{tabular}{|c|ccccc|}\hline
$\alpha$&{$\Bra{1^{+}_\alpha}\O^{\beta^-}_1\Ket{0^{+}}$}
& $\Bra{0^{+}_f}\O^{\beta^-}_1\Ket{1^{+}_\alpha}$
& $\Bra{1^{+}_\alpha}\O^{\beta^-}_1\Ket{0^{+}}\cdot
\Bra{0^{+}_f}\O^{\beta^-}_1\Ket{1^{+}_\alpha}$&${{\cal D}_{1 ^+_\alpha,1}}$
&${\cal M}_{2\nu}^{GT}$ \\
\hline\hline
&&\underline{case (a):}& $v_s=35$, $v_t=65$&&\\
$1$&$2.240$&$0.088$&$0.198$&$3.40$&$0.058$\\
 $2$&$-4.357$&$0.041$&$-0.176$&$11.32$&$-0.016$\\
&total&&&&$0.042$\\
\hline&&\underline{case (b):}& $v_s=40$, $v_t=60$&&\\
$1$&$2.240$&$0.142$&$0.318$&$4.01$&$0.079$\\
 $2$&$-4.357$&$0.002$&$0.010$&$11.91$&$0.001$\\
&total&&&&$0.080$\\
\hline
\end{tabular}
\end{center}
\end{table}

The way in which ${\cal M}_{2\nu}^{GT}(0^+_1)$  are constructed   from individual
transitions to the intermediate states $\ket{1_\alpha^+}$ is shown in Table I. We see  that
the GTR contributes either destructively - as
 in case (a), or its contribution is negligibly small -  as in case (b).
 The first result is consistent with the SM calculation \cite{Zha90}. The interference
 effect is still more pronounced on the GT  strength going to  the ground state
 in $^{48}$Ti. In fact, for all  practical purposes  it turns out to be null in case (a):
 $S^{\beta\beta^-}_{GT}(0^+_1)=(0.198-0.176)^2/3=0.000$, and extremely
small in  case (b): $S^{\beta\beta^-}_{GT}(0^+_1)=0.036$. Note, however, that the quenching
mainly comes from out of phase contributions among the seniority-zero and seniority-four
components in the wave function  \rf{5.10}, which is responsible for relatively small
values of the moments $\Bra{0^{+}_f}\O^{\beta^-}_1\Ket{1^{+}_\alpha}$  shown in Table I.
\begin{table}[h]
\begin{center}
\caption
{Results for  GT, F and total $2\nu\beta\beta$
matrix elements
in units of (MeV)$^{-1}$.  }
\renewcommand{\tabcolsep}{2.pc} 
\renewcommand{\arraystretch}{1.2} 
\label{table2}
\bigskip
\begin{tabular}{|c|ccc|}\hline
case&${\cal M}_{2\nu}^{GT}$&${\cal M}_{2\nu}^{F}$&${\cal M}_{2\nu}$\\
\hline
(a)&$0.043$&$0.008$&$0.035$\\
(b)&$0.080$&$0.011$&$0.069$\\\hline
\end{tabular}
\end{center}
\end{table}

The calculated  GT, F and total $2\nu\beta\beta$  matrix elements are
shown in Table II. One sees that the F moments are quite significant
and that    they  cannot be  neglected as is usually done.
 In point of  fact, in several  calculations, where sizable contribution to the  total
   ${\cal M}_{0\nu}$ moment was found to come from the virtual $0^+$ states, the F contribution
   to the ${\cal M}_{2\nu}$ moment had been omitted. This is  manifestly inconsistent.
\footnote{A more careful study should go beyond the allowed approximation
considered here and include the higher order contributions
in the weak Hamiltonian as well \cite{Bar95, Bar99}.}

Similarly, the model  wave function for the excited state
$\ket{0^+_2}$ in $^{48}Ti$ is basically  composed  of seniority-four
basis states, \ie
\br
\ket{0^+_2}&=&
\left.\begin{array}{ll} 0.430\\0.504 \end{array}\right\}
 \ket{f^2_{7/2},f^2_{7/2}; 2}
-\left.\begin{array}{ll} 0.781 \\0.760 \end{array}\right\}
 \ket{f^2_{7/2},f^2_{7/2};  4}
+\left.\begin{array}{ll} 0.373 \\0.307 \end{array}\right\}
 \ket{f^2_{7/2},f^2_{7/2}; 6}+\cdots.
\nn\\
\label{5.11}\er
Unlike for the ground state, in case (a) appears an appreciable amount of the
GT strength, $S^{\beta\beta^-}_{GT}(0^+_2)=0.170$, in the state $\ket{0^+_2}$.
Nevertheless, we will not discuss here the $\beta\beta$-decay rate for the exited ${0^+}$
state because
the model does not reproduce satisfactorily the excitation energy;
namely, we get $5.72$ MeV,  while the experimental value is $3.00$ MeV.
Finally, we note that in  case (b)
$S^{\beta\beta^-}_{GT}(0^+_2)=0.038$.

\section{Concluding Remarks}

The QTDA  does not suffer from
the inconveniences that  have been listed in Section III in relation
to the QRPA.
More specifically, the similarities and the dissimilarities
between the two models are:
\bnu
\item
While the QTDA  contains two ``$\beta^-$- like''
vertices (see Fig. 3), in the QRPA  we  always
approximate  one of  them
by a ``$\beta^+$- like'' vertex (see Fig. 2).
This statement  is valid for  all variations of the
standard QRPA, such as the TVQRPA, PQRTA, RQRPA and SCQRPA.

\item
 The  QTDA moments $\Bra{\lambda^+_\alpha}\O^{\beta^-}_\lambda\Ket{ 0^{+}}$,
given by \rf{4.7}, produce  the F and GT resonances
\cite{Nak82}, in the same manner as in the QRPA.
Furthermore, as the  backward-going QRPA contributions  have rather
little impact on  the ``$\beta^-$- like'' transition strength, given
by the first
equation in  \rf{3.5}, both models yield  very similar results.
 \item
Similarly, moments $\Bra{ 0^{+}_1}\O^{\beta^-}_\lambda \Ket{\lambda^{+}_\alpha}$,
given by \rf{4.8}, are strongly reduced by the residual interaction, as are the
QRPA moments $\Bra{\lambda^+_\alpha}\O^{\beta^+}_\lambda\Ket{ 0^{+}}$, given by the
second equation in  \rf{3.5},  restoring in this way   the isospin SU(2) and
Wigner SU(4) symmetries,  broken initially by the mean field. We know that in QRPA
this symmetry-reestablishment  takes place   through the cancellation effect between
the forward and the backward going contributions. In fact, in several works
\cite {Hir90, Krm94, Krm94a, Bar99} we have used the property of maximal restoration of
the SU(4) symmetry to fix the value of the $pp$ parameter $t$. Instead, in the  QTDA  the
quenching comes from out of phase contributions among  seniority-zero and seniority-four
configurations in the wave function of the final state $\ket{ 0^{+}_1}$.
\item
In QTDA, unlike in QRPA, the  intermediate states $\ket{\lambda^+_\alpha}$  have a unique meaning.
\item
The QTDA never collapses, and therefore  the  amplitude
${\cal M}_{2{\nu}}$ does not have a pole anymore, as happens in
\rf{3.24}.
 So, one can expect that, as far as model parameters are concerned  ${\cal M}_{2{\nu}}$
 will behave more moderately in the QTDA than in the QRPA.
\item
In the QTDA one obtains the transition amplitudes  ${\cal M}_{2{\nu}}$ for the excited $0^+_f$ states
 without any additional effort. One can also easily evaluate the decays to the  excited final
 $2^+_f$ states; it is sufficient to diagonalize  the Hamiltonian \rf{4.1} in the
 $\ket{(p_1p_2)J_p,(n_1n_2)J_n;2^+}$ basis, and to calculate the matching transition operator \rf{4.8}.
\item
In the QTDA the energy distributions of the double GT transition strengths  \rf{2.7} can
be evaluated directly and, the corresponding sum rule, given by \rf{2.12}, could
 be occasionally violated to some extent. Contrarily, the Ikeda sum rule,
 given by \rf{2.6}, is always fully conserved in this model.
 \enu

The remarks quoted above suggest that perhaps the QTDA  might be a more "natural" and a more
suitable nuclear structure framework for describing   the $\beta\beta$-decay than the QRPA.
In fact, one should mention again that the QRPA was originally  formulated for the single
$\beta^\pm$-decays \cite{Hal67}, and  only later adapted for the $\beta\beta$-decays
via the M1 and M2 \cite{Vog86,Civ87,Tom87,Eng88} ansatz. The TVQRPA, which was
mathematically tailored, specifically  for the $\beta\beta$-decay, is free of these
averaging procedures  and has the correct BCS limit given by \rf{3.21}. Nevertheless the
latest model has received rather little attention  in the literature.  What's more, quite
recently it has  been  claimed  that the overlap \rf{3.9} gives rise to  an additional
suppression mechanism for the   NME's ${\cal M}_{2\nu}$ and ${\cal M}_{0\nu}$ \cite{Sim04, Alv04}.
 We fully disagree with such a view.

We do not suggest that the QRPA should be substituted by the QTDA. We merely state that
the use of these two nuclear models in a joint manner  should very likely  reduce the
uncertainties in the  evaluation of the NME's, which is at present one of the principal
 worries  in the nuclear physics   community,  and which   has engendered a great deal
 of activity in recent years  \cite{Civ03,Rod03,Suh05,Rod05}. One cannot but highlight
 the Suhonen's article \cite{Suh05} where he argues that within the QRPA it is not
 possible to account simultaneously, \ie with the same set of model parameters,
 for the simple and double $\beta$-decays.

In summary, in this work we have demonstrated that  the simple version of the QTDA
proposed here is able to account for the suppression of   ${\cal M}_{2\nu}$, which
was  the major merit of the QRPA. Moreover, we feel that  this model comprises all
essential nuclear structure  ingredients that are needed for describing the
$\beta\beta$-decay processes. Of course, whether this is totally or only partly
true has still to be tested, and it might  be convenient to consider  some additional
refinements in the future.    Their incorporation, however,  should  not, in principle,
involve  serious difficulties. For instance, the number projection procedure can be easily
implemented  whenever required \cite{Krm93}.  Regarding this issue we are convinced
that,  despite the  present-day  lack of consensus among nuclear theorists on how  to
derive the NME's   in a direct and controlled manner,  they  will be able to surmount this
obstacle  in the near future  without  having  to resort to extremely complicated theories.
In fact, nuclear physics is not merely complicated    mathematics: it requires much art
to discover the most important degrees of freedom and to disentangle the underlying
symmetries, for the purpose of building very
imaginative  models, and to skilfully manipulate  the pertinent   adjustable parameters.
 {One should always keep in mind  Milton's witty remark: {\em   The very essence of
 truth is  plainness and brightness.} Very likely, once fixed
 the NME's, it will be possible to answer some fundamental questions about neutrinos.

I'm very  sure Dubravko would agree with all that has been said above, and this greatly
encourages me to pursue the presented  line of research and it is a further reason to
dedicate this article to him.

\begin{center}
{\bf ACKNOWLEDGEMENTS}
\end{center}
I am thankful to Arturo Samana for doing  the drawings and for the  help
in numerical calculations.
I am also extremely grateful to Gordana
Tadi\'c for her very careful reading of the manuscript and for many
very useful suggestions.


\end{document}